*Selection on moral hazard in the Swiss market for mandatory health insurance: Empirical evidence from Swiss Household Panel data*

Igor N. Francetic

This version: September 23

**Abstract**

Selection on moral hazard represents the tendency to select a specific health insurance coverage depending on the heterogeneity in utilisation "slopes". I use data from the Swiss Household Panel and from publicly available regulatory data to explore the extent of selection on slopes in the Swiss managed competition system. I estimate responses in terms of (log) doctor visits to lowest and highest deductible levels using Roy-type models, identifying marginal treatment effects with local instrumental variables. The response to high coverage plans (i.e. plans with the lowest deductible level) among high moral hazard types is 25-35 percent higher than average.

**Author affiliation and contacts**: Health Organization, Policy and Economics (HOPE) Group, Centre for Primary Care and Health Services Research, School of Health Sciences, University of Manchester, Oxford Road, Manchester, United Kingdom. Contact: igor.francetic@manchester.ac.uk. ORCID: 0000-0002-7423-0090.

**Conflict of interest disclosure**: None.

**Funding**: This research did not receive any specific grant from funding agencies in the public, commercial, or not-for-profit sectors.

**Data availability statement**: The main data that support the findings of this study are available from the Swiss Centre of Expertise in the Social Sciences. Restrictions apply to the availability of these data, which were used under license for this study. Data are available at https://forscenter.ch/projects/swiss-household-panel/ with the permission of the owner. Part of the data supporting the findings were derived from reports published by the Swiss Federal Office of Public health and available in the public domain: https://www.bag.admin.ch/bag/en/home/zahlen-und-statistiken/statistiken-zur-krankenversicherung.html. A replication packages including software codes and aggregated data from publicly available sources will be available at https://github.com/igorfrancetic/.

**Acknowledgments**: This study has been realised using data collected by the Swiss Household Panel (SHP), which is based at the Swiss Centre of Expertise in the Social Sciences FORS. The project is supported by the Swiss National Science Foundation. The author is grateful to Prof. Carlo De Pietro, Dr. Luke Munford, Dr. Anna Wilding, Prof. Heather Brown, Hyacinthe Müller, Prof. Günther Fink, two anonymous reviewers, participants to the 2021 Congress of the Scottish Economic Society, Winter 2022 Health Economists' Study group at the University of Leeds, 2022 Swiss Health Economics Workshop 2022 at the CSS Institute in Luzern, and the 2022 American-European Health Economics Study Group at UPF in Barcelona for useful feedback on previous versions on the manuscript. All errors are my own.

# 1. Introduction

Ever since the seminal contribution by Arrow (1963), moral hazard in health insurance has been a lively area of research in health economics. In a nutshell, moral hazard emerges as higher insurance coverage generates an increased use of healthcare, regardless of actual needs (Zweifel and Manning 2000). This phenomenon has been studied widely across different high-income settings, for example the U.S. (Einav and Finkelstein 2018; Pauly 2004), the Netherlands (Alessie et al. 2020), France (Sevilla-Dedieu, Billaudeau, and Paraponaris 2020) and Germany (Thönnes 2019). A related concept is that of adverse selection, i.e. the idea that individuals with higher expected health expenditures self-select into plans with higher insurance coverage. Adverse selection implicitly suggests that individuals with high and low health insurance coverage may be intrinsically different, complicating any trivial assessment of the extent of moral hazard. (Cutler and Reber 1998; Chiappori and Salanie 2000; Cabral 2017; Olivella and Vera-Hernández 2013). A common feature of health insurance plans regarded as a signal of adverse selection is the endogenous choice of cost-sharing models (Becker and Zweifel 2005).

Einav and Finkelstein (2018) provide a detailed review of the theoretical and empirical evidence on moral hazard in insurance markets. From an individual perspective, health insurance allows the insured to smooth consumption across health states, transferring consumption from periods where they are healthy to periods when they are sick (or from states of low to states of high marginal utility of income). An optimal insurance contract would ensure equal marginal utility in all states. However, such first best contracts are not feasible under moral hazard due to the not contractible nature of health status, which alongside adverse selection generates the well-known positive correlation between health insurance coverage and healthcare utilisation (Einav and Finkelstein 2011). The main (second best) solution to overcome this correlation has been the introduction of cost sharing schemes; deductibles and co-insurance rates introduce a trade-off between additional spending due to moral hazard and providing insurance coverage. From a health systems' perspective, overlooking numerous important implementation details and abstracting from its pure financing role, mandatory health insurance represents a mechanism of solidarity between healthy and sick individuals (McGuire and van Kleef 2018). In countries that decided to enforce a health insurance mandate, cost sharing rules shielding the system from overconsumption due to moral hazard are necessarily applied indiscriminately to populations (or groups thereof, e.g. no deductibles for children but the same deductibles and co-payments for all adults). In settings where premiums are unrelated to income, this can introduce discriminatory distortion, penalising low income and sick individuals. Hence, measuring the extent of moral hazard is of crucial for health policy makers to understand the extent to which cost sharing arrangements are effective in limiting the efficiency losses due to moral hazard whilst maintaining risk sharing and solidarity (D. Powell and Goldman 2021).

An interesting notion bridging adverse selection and moral hazard is the idea of selection on moral hazard. Einav et al. (2013) explain that selection on moral hazard is driven by the slope of spending: the incremental healthcare use due to a more comprehensive coverage. So, in the decision to select a specific coverage (including the level of cost-sharing), individuals incorporate their anticipated behavioural responses to health insurance, as well as their risk preferences and risk profile. This component of adverse selection is distinct from the traditional idea of (adverse) selection on levels of expected health risk, mentioned above. Crucially, the phenomenon of selection on moral hazard implies that individuals have heterogeneous responsiveness to health insurance coverage. High-responsiveness individuals (i.e. high moral hazard types) will then be more likely to purchase more comprehensive health insurance coverage because they anticipate the higher healthcare utilisation due to the coverage itself. An understanding of the extent of selection on moral hazard is again very relevant for countries with health systems funded via mandatory health insurance as it allows to gauge the full implications of cost sharing rules and premium increases for healthcare utilisation. This is particularly important given that individuals with different responsiveness to coverage face the same cost sharing rules, implicitly calibrated to an average level of moral hazard in the population.

My contribution to the literature is twofold. Firstly, I add to the mixed body of evidence assessing the existence and extent of selection on moral hazard in health insurance markets. Secondly, I contribute to the understanding of patient responses to mandatory health insurance in a setting – Switzerland - characterised by a highly regulated health insurance market, constant and sharp increases in healthcare expenditure well above GDP growth, and a related (cost driven) growth in health insurance premiums. Whilst these latter findings are undeniably country-specific, they can be of interest for policy makers in countries with analogous health insurance mandates, such as Germany or the Netherlands.

The extent of selection on moral hazard in health insurance markets remains an open empirical question. For health insurance plan choices of large employer in the U.S., Einav et al. (2013) found that selection on heterogeneous moral hazard was significant, but that overall had mild welfare consequences. Similarly, Péron and Dormont (2018) find substantial selection effects on heterogeneous moral hazard in the market for voluntary supplementary insurance in France. On the other hand, Alessie et al. (2020) found no evidence of selection on moral hazard using survey data from the Netherlands in the context of voluntary deductible choices. In the case of the mandatory health insurance in Switzerland, results from previous studies seem to favour the hypothesis that higher insurance deductibles reduce healthcare consumption via a reduction in the overall moral hazard effect (Boes and Gerfin 2016; Gerfin and Schellhorn 2006). More recently, Zabrodina (2022) tested the extent of so-called "timing moral hazard". The author finds sizeable behavioural responses to insurance coverage among ex-ante comparable insured individuals of a large Swiss insurer who are exposed to quasi-random variation in the timing of health shocks and - to reduce their out-of-pocket payments in the upcoming year - strategically modulate healthcare use depending on how far the shock realisation is from reset of deductibles at the end of the calendar year. The recent paper by Zabrodina (2022) complements well previous work by Cabral (2017), who exploits claim timing to identify instances of timing moral hazard that give rise to ex post adverse selection, specifically in relation to accumulated delayed non-urgent treatments. On the other hand, Schellhorn (2001) suggested that reductions in healthcare utilisation associated with higher deductible in Switzerland may simply reflect self-selection rather than behavioural responses related to moral hazard. To the best of my knowledge, this is the first study measuring the extent of selection on moral hazard using data for a representative sample of Swiss residents.

My study uses a mix of survey data from the Swiss Household Panel and publicly available information from the health insurance markets regulator. I estimate a Roy-type model allowing for selection on gains from coverage, using a local IV estimator relying on local average premiums and coverage with supplementary health insurance as excluded instruments. I estimate the effect of coverage on the (log) number of doctor visits, modelling the selection of the lowest and respectively the highest deductible levels. I report the marginal response to coverage along the distribution of the unobserved resistance to select the specific coverage level, which can be interpreted as the inverse of the multiplicative effect consistent with moral hazard. My findings suggest that individuals with the lowest resistance to selecting a low deductible (i.e. high moral hazard types) have a substantially and significantly higher healthcare utilisation response (in terms of doctor visits), a finding consistent with selection on moral hazard. Parametric and semiparametric estimators both suggest that the response to the highest coverage plans (i.e. with lowest deductibles) among high moral hazard types is 25-35 percent higher than average. The finding is robust to restricting the analysis to individuals with stable individual characteristics, hence reducing concerns of confounding due to changes in personal circumstances.

The rest of the paper is structured as follows: Section 2 describes the Swiss health insurance market and institutional setting, Section 3 outlines a model which provides the main testable hypotheses. Section 4 discusses the empirical approach and the main sources of data. Finally, Section 5 presents the main results, which I discuss in Section 6.

**2. Setting**

*2.1 Statutory health insurance in Switzerland*

With about 8.5 million people located in the heart of Europe, Switzerland is a confederation of 26 independent Cantons responsible for steering their health systems. Switzerland ranks fourth among OECD countries in terms of GDP per capita, with 71 thousand USD for 2019 (OECD 2020a), whilst total life expectancy at birth was 83.8 in 2018 (OECD 2020c). Switzerland also reports consistently high levels of health spending per capita, reaching 7'732 USD ppp[1] in 2019 (OECD 2020b). Since 1996, Switzerland has organized healthcare financing with a managed competition system (Crivelli 2020). All Swiss residents are required to obtain health insurance coverage from a number of insurers offering standardized plans in a strictly regulated marketplace, akin to the one set up in the U.S. with the Affordable Care Act (Courtemanche, Marton, and Yelowitz 2016). Insurers on the market offer a set of plans which vary along two main dimensions: yearly deductible level and managed care option. Options for yearly deductible are 300, 500, 1'000, 1'500, 2'000 and 2'500 Swiss francs. Higher deductibles are associated to premium reductions up to 70 percent. Managed care plans (e.g. family doctor, gatekeeper through phone consultation with a medical consultant or HMO) also offer premiums up to 20 percent cheaper compared to the basic plan featuring free provider choice. The benefit package is standard across all plans and everyone is subject to a 10 percent co-payment for all costs in excess of the deductible up to a 700 Swiss francs cap. The marketplace is further regulated with a risk-adjustment in cost differences between models with free provider choice and managed care, as well as a community-rated adjustment on a cantonal level based on age, gender, inpatient care and consumption of medicines in the previous year (Kaufmann, Schmid, and Boes 2017; Federal Office of Public Health 2020a). Premium discrimination is also allowed across three age groups: children (0-18 years), young adults (18-25) and adults (26 and older). Thus, individuals in the same Canton and age group face the same healthcare plan choice set, irrespective of income. To correct this strong element of inequity, besides broader social security mechanisms, Cantons grant earmarked means-tested subsidies to about 30 percent of residents based on taxable income in the previous years (Kaufmann, Schmid, and Boes 2017). Furthermore, healthcare costs for inpatient care are partially (55 percent) financed through general taxation by Cantons.

Assuming rational responses to price differences for equivalent plans, competition among insurers should lead to efficient premiums and high quality of service. However, switching rates have been historically quite low (Laske-Aldershof et al. 2004), likely reflecting behavioural biases (Krieger and Felder 2013; Frank and Lamiraud 2009) and bundling with supplementary (voluntary) health insurance (Dormont, Geoffard, and Lamiraud 2009). Heterogeneity in the scale of insurers and other market inefficiencies generate wide differences in premiums and – mostly due to inertia - citizens often forgo savings up to 40 percent for a virtually identical coverage (Crivelli 2020). Switching rates have slightly increased in the last years, possibly in response to the steady growth of premiums parallel to the rise in health expenditures (Crivelli 2020) and to increased ease of access to online tools to compare premiums and facilitate switching (Lako, Rosenau, and Daw 2011). Additionally, the strict regulation of standardized health insurance plans entails – by design – a set of dominated (non-optimal) plans offered on the market, which result in further welfare losses (Biener and Zou 2021).

*2.2 Deductible choice and potential role of selection on moral hazard in the Swiss setting*

Every year, by the end of November, Swiss residents can notify to insurance companies their decision to change insurer and/or coverage for the subsequent year in relation to mandatory health insurance. Insurers cannot refuse coverage to any individual in relation to mandatory health insurance, whilst they are free to refuse coverage and conduct risk assessments for voluntary health insurance plans, which are subject to private insurance laws. In the regulated setting outlined above, consumers decide to change insurer or coverage in response to premium levels or shocks to individual characteristics, considering both private (socioeconomic status, risk factors, risk preferences, healthcare consumption patterns, health status, etc.) and public information (i.e. new premiums announced every September for the coming year).

---

[1] OECD used the purchasing power adjustment based on most recent Actual Individual Consumption, or AIC (OECD 2019).

On the one hand, the traditional concept of "ex post" moral hazard in health insurance suggests that individuals with higher coverage – facing milder incentives to limit their healthcare consumption – tend to have higher levels of healthcare utilisation. This effect has typically been looked in terms of an average tendency to use more medical care, mostly overlooking the heterogeneity in this responsiveness to coverage. On the other hand, adverse selection suggests that individuals with high expected healthcare costs sort themselves into plans with high coverage, i.e. low deductible. Therefore, without underlying changes in risk type and assuming constant preferences (at least in the short run, from one year to the other), sharp changes in the level of deductible (high to low, or low to high) are consistent with optimal responses to changes in:

- Premiums (i.e. relative prices of different plans);
- Household's budget constraint (e.g. income reductions that lead families to choose higher deductibles, in order to pay lower premiums);
- Anticipated healthcare consumption intentions;
- Anticipated behavioural responses to different coverage levels.

The notion of "selection on moral hazard" is consistent with this last point. In the spirit of Einav et al. (2013), when deciding their coverage for year *t+1*, consumer choice reflect heterogeneous responses to health insurance coverage in *t+1*. Intuitively moral hazard materializes as a tendency to switch to higher coverage (i.e. lower deductible) – from a previous lower coverage level - for individuals expecting to have a stronger response to coverage in terms healthcare utilisation when faced with a healthcare need. The opposite should also be true, namely that switching to lower from a previously higher coverage should be associated with a milder response to the health shock.

## 3. A simple model of selection on moral hazard

To build the intuition of my main research hypothesis, this section outlines a simple two-period model that draws heavily from Einav et al. (2013). This stylized decision process is fully consistent with the Swiss institutional background and health insurance choices described above.

In the first period, the utility-maximising risk-averse agent makes a financial decision, selecting her optimal coverage level for the second period, anticipating her optimal response in terms of healthcare utilisation to her future healthcare needs. In the second period, given a coverage level and a realized healthcare need, she decides her optimal level of healthcare utilisation, trading off health and money.

Echoing the original contribution (Einav et al. 2013), I assume the period 2 utility to be additively separable in health and money. The quadratic (monetized) health utility component is given by

$$H[m - \lambda^t; \omega] = (m - \lambda^t) - \frac{1}{2\omega_i}(m - \lambda^t)^2 \qquad (1)$$

That is, the agent has concave preferences in the net benefit of treatment $m$ given her healthcare need type $\lambda^{t \in (High, Low)}$. Utility increases with $(m - \lambda^t)$ until a switching point where cost of treatment outweighs the benefits, and utility starts to decrease. The switching point is higher for high healthcare need types, i.e. sicker individuals with higher need for (monetized) healthcare ($\lambda^{High} > \lambda^{Low}$). In this parameterization, $\omega$ shapes the curvature of the relationship and will be interpreted as a coefficient of moral hazard, shifts the level of optimal spending.

The monetary component of utility simply given by

$$Y[m; c_i, \pi(c_i)] = y - c_i m - \pi(1 - c_i) \quad (2)$$

The agent's period income $y$ is given. To simplify matters, I characterize the insurance contract as a combination of a co-payment rate $c_i \in [0,1]$ and corresponding premium defined by a tariff $\pi(1 - c_i)$, which is a monotonic decreasing function ($\pi'(1 - c_i) < 0$) of the share of healthcare costs covered by the insurer $(1 - c_i)$. At the time of the utilisation decision, $\lambda^t$, $\omega$, $c_i$, and $\pi(1 - c_i)$ are known. Hence, the agent maximises her period utility choosing

$$\max_m u[m; \lambda^t, c_i, \pi(1 - c_i)] = (m - \lambda^t) - \frac{1}{2\omega}(m - \lambda^t)^2 + y - c_i m - \pi(1 - c_i) \quad (3)$$

The optimal utilisation level is given by

$$m^* = \omega(1 - c_i) + \lambda^t \quad (4)$$

Equation (4) suggests that the optimal utilisation level increases with moral hazard, the share of costs covered by the insurer, and the realized healthcare need. Moving from a lower to a higher level of coverage (from high to low $c_i$) should result in a higher discretionary healthcare utilisation for the same level of non-discretionary healthcare need.

In period 1, the agent anticipates her optimal period 2 utilisation response $m^*$, conditional on her realized healthcare need and her coverage choice. I assume that the agent assigns a probability $p$ to a realization $\lambda^{High}$, and probability $(1 - p)$ to realization $\lambda^{Low}$. Given the (*ex-ante*) financial nature of health insurance contracts, I model her period 1 choice as a decision over future financial outcomes (i.e. the monetary component of the period utility). Assuming a simple logarithmic utility function to capture risk-averse preferences, the agent chooses $c_i$ in order to maximise her expected utility, as follows

$$\max_c E\{u(c, \pi(1 - c_i); y, \omega, \lambda^t)\} =$$
$$p \ln[y - c_i(\omega(1 - c_i) + \lambda^{High}) - \pi(1 - c_i)] + (1 - p)\ln[y - c_i(\omega(1 - c_i) + \lambda^{Low}) - \pi(1 - c_i)]$$

After rearranging terms, the optimal level of co-payment can be expressed as

$$c^* = \frac{1}{2} + \frac{1}{2\omega}[E[\lambda] - \pi'(1 - c)] \quad (5)$$

Appendix A provides full details regarding the model solution. The central result in expression (5) is that higher value of moral hazard $\omega$ are associated with higher coverage (i.e. lower values of co-payment). The model also suggests that a higher expected healthcare need is associated with higher coverage, consistently with a classic adverse selection argument. The effect of $\pi'(1 - c)$ (which is negative) captures the influence of tariff function steepness on coverage choice $c^*$.

## 4. Data and Methods

*4.1 Data*

The main data source for this work is the Swiss Household Panel, henceforth SHP (Tillmann et al. 2022), a nationally representative longitudinal study running since 1999. Three waves collected prior to the Covid-19 pandemic in 2017/2018, 2018/2019 and 2019/2020[2] included data on health insurance coverage choices, except for insurer and premium paid. All waves also include information on a limited number of healthcare utilisation variables and. Alongside SHP, I use data on market premiums published by the regulatory authority for the mandatory health insurance in Switzerland. A full list of data sources is provided in Appendix B.1.

---
[2] Data for the 2019/2020 wave were collected later than February 2020, whilst the first ever Covid-19 case in Switzerland were recorded at the end of the February 2020.

To define my core analytical sample, I apply a few sample restrictions to the raw SPH data. First, to focus as much as possible on financially independent individuals and avoid mixing two different premium classes (see Section 2.1), I focus only on participants aged 26 or older. Second, I necessarily focus on respondents reporting responses on the key questions about deductible level and type of health insurance plan. Third, since my study is based on responses capturing insurance coverage choices for the 2017-2019 waves, when studying changes in health insurance coverage from one wave to another I necessarily restrict my analysis to years 2018 and 2019. Finally, my analytical sample is slightly restricted by some limited patterns of missing information on some of covariates, determining a slight variation in the number of observations across years. In Appendix B.2 I detail the construction of the analytical samples reporting numbers of observations at each of these steps.

*4. 2. Main outcome*

My main proxy outcome for healthcare utilisation is the number of doctor visits in a given calendar year. The question captures visits to specialist and general practitioners in outpatient settings, explicitly excluding dentists, any type of hospital care, and other types of healthcare providers (mental health, physiotherapy, etc.).

Deciding whether to visit a doctor in the first place and deciding how many times to visit are two inherently different decisions, possibly also faced by different sub-populations. Figure 1 shows that in my final sample, 78.4 percent of survey respondents reported having at least one doctor visit. Among these, the median number of visits was 3. Individuals reporting no doctor visit in the previous 12 months are markedly different in their observable characteristics compared to those with one or more doctor visits, suggesting that they are also likely to differ in unobservable characteristics (see Appendix B). Since I am not able to disentangle moral hazard from these unobservable factors, I focus decision on the intensive margin of healthcare utilisation. I use the (log) number of doctor visits in a given calendar year conditional on having at least 1.

Whilst the choice of this outcomes was primarily driven by data, as this was the only consistently recorded measure of healthcare utilisation available in the SHP, it is also commonly used in the relevant literature. Examples include previous work on the role of insurance in healthcare utilisation in Switzerland (Gerfin and Schellhorn 2006), and studies based on SHARE or Swiss Health Survey data exploring socioeconomic inequalities in healthcare use across different countries (Allin, Masseria, and Mossialos 2009; Kalouguina and Wagner 2020).

Beyond previous literature, looking at the number of visits with a family doctor to proxy healthcare utilisation in the context of the Swiss health system is also interesting for at least three contextual policy reasons. Firstly, family doctors remain the first point of contact with the health system in Switzerland, referring patient to subsequent inpatient care and triggering purchases of prescription drugs. Secondly, visits to family doctors represent the largest share of expenditures paid by mandatory health insurance funds (consistently about 23% overall), closely followed by medicines (Federal Office of Public Health 2022). Relatedly, contrary to inpatient care, payments towards outpatient care are not partially subsidised by cantons, implying a more salient link between use and insurance coverage for this class of healthcare treatments. Thirdly, Swiss authorities directly regulate the supply of family doctors by imposing a moratorium on new doctor practices in a bid to control healthcare costs (Fuino, Trein, and Wagner 2022).

Figure 1: Distribution of doctor visits on the extensive and intensive margins

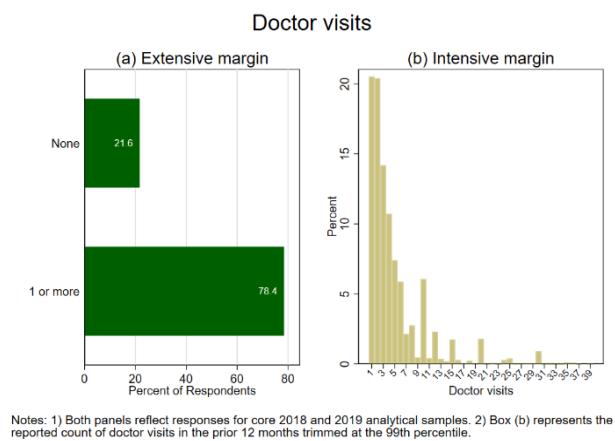

### 4.3. Main exposure

From the SHP I also obtain the level of deductible in each year measured in Swiss francs, as well as the type of plan (i.e. free choice of provider, or managed care options). My focus in on the choice deductible, which effectively determines the level of co-payment faced by individuals for their yearly healthcare expenditures financed through the mandatory health insurance system (beyond the 10 percent co-payment faced by anyone for all expenses beyond the capped deductible, see Section 2 for more details). Figure 2 represents the distribution of deductible choices in 2018 and 2019, suggesting substantial stability across the two years considered. In my main empirical approach I focus on the choice of either of the two extreme options (300 and 2'500 Swiss francs), as these are both the most common and generally rational options whilst the intermediate choices are generally dominated (Biener and Zou 2021). In my sample, about 81 percent of respondents do not switch their deductible level for years 2018 (compared to 2017) and 2019 (compared to 2018).

Figure 2: Distribution of deductible choices in 2018 and 2019

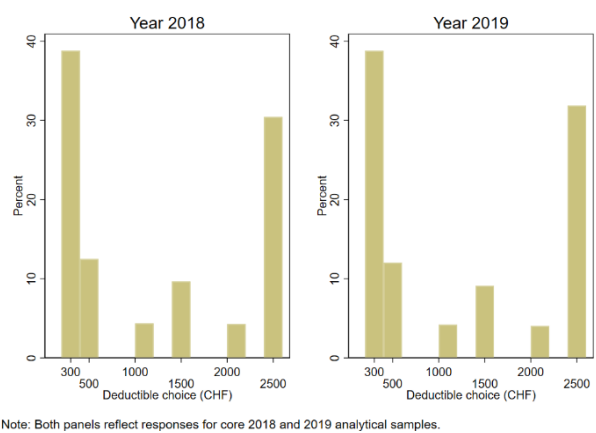

To simplify the patterns of deductible switching behaviours observed in the data, Table 1 summarizes 5 mutually exclusive groups. I discuss the extent to which these groups are comparable and the corresponding implications for my empirical approach below.

Table 1: Deductible switching groups in the core 2018 and 2019 samples

| Groups | Switching behaviour | Proportion Year 2018 (vs. 2017) | Proportion Year 2019 (vs. 2018) |
|---|---|---|---|
| No switch | Individuals reporting the same deductible level in years $t$ and $t-1$ | 0.813 | 0.820 |
| Strong drop | Large reductions of 1'000 – 2'200 Swiss francs, moving highest deductible levels in $t-1$ to the lowest in $t$, likely to reflect conscious changes to coverage | 0.041 | 0.042 |
| Mild drop | Small reductions (200-700 Swiss francs) in deductible levels, likely to reflect small adjustments to accommodate premium differences from year to year and budget constraints | 0.050 | 0.051 |
| Mild increase | Small increases (200-700 Swiss francs) in deductible levels, likely to reflect small adjustments to accommodate premium differences from year to year and budget constraints | 0.050 | 0.048 |
| Strong increase | Large increases of 1'000 – 2'200 Swiss francs, moving from lowest deductible levels in $t-1$ to the highest in $t$, likely to reflect conscious changes to coverage | 0.046 | 0.039 |
| | Observations | 4087 | 4074 |

*4. 4. Data on health insurance premiums*

To complement SHP data, I use publicly available data on premiums published by the oversight authority of the Swiss mandatory health insurance market (Federal Office of Public Health 2020b). These data offer a complete overview of the supply of mandatory health insurance plans across Switzerland.

The SHP data unfortunately does not report the level of premiums faced by individuals. To recover a measure of the price information available to individuals when making choices regarding their health insurance coverage for the following year, I construct the average market premium for the type of coverage purchased. Specifically, for any given reported coverage choice (represented by combinations of age group, deductible, type of plan and canton of residence, see Section 2), I generate the average premium in the relevant market. Appendix B.3 shows that the distribution of these locally defined premiums are generally precise enough to represent a meaningful price signal for consumers, with relatively little dispersion around the mean. Crucially, these price signals are exogenous for households (i.e. households have no sense of how their health insurance premium will change in the next year, and have no way to directly influence it). Households then respond to premium announcements in September to make decisions by the end of November about their health insurance coverage from the next January onwards.

*4.5. Other variables*

The SHP includes a wide range of personal and household characteristics. Besides standard socio-economic characteristics, these include self-reported receipt of means-tested subsidies to cover part of the mandatory health insurance premium, and coverage with a supplementary health insurance plans.

The data also identifies the canton where respondents were living at the time of the survey. Albeit not perfectly, controlling for canton fixed effects in the analysis allows to account for structural differences in costs of healthcare and differences in supply-side availability.

The survey collects detailed information about the health status of respondents. These variables include self-reported health status, reporting a chronic condition, smoking behaviour, and an index measuring the respondent's need for medications in everyday life (0= no needs, 10= very high needs). The survey also reports the occurrence of illnesses, accidents or serious health problems in the survey year. I use this variable to construct an indicator of health shock (akin to the "surprise in healthcare need" defined in Appendix A) defined as a new illness, accident or serious health problem in year $t$ (i.e. with no occurrence reported in year $t-1$).

Having only three waves of data on insurance choices available and using lagged values to construct both the exposure and some control variables, I am left with two waves of data for my analysis, namely 2018 and 2019. Table 2 below the characteristics for individuals in the 2019 analytical sample, overall and separately across the exposure groups used in the main analysis, that is those who selected a deductible of either 300 or 2'500 Swiss francs.

Table 2: Descriptive statistics for individuals in the 2019 analytical sample, overall and for low and high deductibles

|  | 2019 | | Low deductible 300 Swiss francs | | High deductible 2,500 Swiss francs | |
|---|---|---|---|---|---|---|
| **Variable** | **Mean** | **Std. Dev.** | **Mean** | **Std. Dev.** | **Mean** | **Std. Dev.** |
| Doctor visits | 5.475 | 9.344 | 6.924 | 12.498 | 3.547 | 4.083 |
| Doctor visits in the previous year | 5.749 | 8.207 | 7.090 | 9.754 | 3.464 | 4.121 |
| Average premium for coverage | 387.709 | 79.348 | 432.254 | 57.369 | 306.900 | 53.087 |
| Supplementary health insurance | 0.340 | | 0.319 | | 0.329 | |
| Health status: Very well | 0.176 | | 0.124 | | 0.255 | |
| Health status: Well | 0.644 | | 0.630 | | 0.672 | |
| Health status: Average | 0.158 | | 0.217 | | 0.067 | |
| Health status: Not very well | 0.018 | | 0.027 | | 0.004 | |
| Health status: Not well at all | 0.003 | | 0.003 | | 0.002 | |
| Health shock (new illness, accident, diseases) | 0.116 | | 0.118 | | 0.117 | |
| Reported smoking in previous year | 0.171 | | 0.175 | | 0.149 | |
| Reported a chronic condition in previous year | 0.461 | | 0.589 | | 0.286 | |
| Reported level of physical activity in previous year | 0.844 | | 0.822 | | 0.893 | |
| Need of medication for everyday life in prev. year | 3.044 | 3.537 | 4.222 | 3.703 | 1.206 | 2.338 |
| Woman | 0.554 | | 0.592 | | 0.503 | |
| Age | 57.502 | 15.169 | 61.462 | 14.499 | 49.950 | 13.892 |
| Years of education | 14.471 | 3.361 | 13.943 | 3.330 | 15.590 | 3.258 |
| Actively occupied | 0.630 | | 0.506 | | 0.826 | |
| Unemployed | 0.008 | | 0.009 | | 0.004 | |
| Not in the labour force | 0.362 | | 0.484 | | 0.170 | |
| Subsidy for mandatory health insurance premium | 0.151 | | 0.160 | | 0.152 | |
| Income per household member | 60581 | 50525 | 55607 | 34381 | 65052 | 43259 |
| Number of people in the household | 2.427 | 1.201 | 2.220 | 1.084 | 2.806 | 1.312 |
| Observations | 4074 | | 1801 | | 1082 | |

Trivially, individuals who chose the lowest level of deductible reported more doctor visits. The majority of people (82 percent) reported being well or very well in the year prior to the interview date. Somewhat accordingly, the share of people reporting weekly physical activity is high (84 percent) and there seems to be a generally limited dependency on drugs in everyday life (3 out of 10). Nevertheless, 46.1 percent of people reported some chronic condition and 17.1 reported smoking in the previous year.

In line with the idea of adverse selection, compared to low deductible types, high deductible individuals are on average older (often retirees, out of the labour force), more likely to report a chronic condition and a higher consumption of medicines. Interestingly, individuals selecting the highest deductible level are also more educated, more likely to be men, actively employed, and report a higher income compared to those selecting the lowest level of deductible. In 2019 sample, about 42.8 percent reported having the lowest level of deductible, whilst 26.7 percent opted for the highest. Between 2018 and 2019, only 18 percent of the individuals in the sample switched deductible: switching patterns are roughly equally distributed across the groups defined in Table 1. In Appendix B.4 I include a similar table for the 2018 analytical sample showing comparable patterns.

## 5. Empirical approach

*5.1. Heterogeneous effects of coverage allowing for selection on unobservables*

My main empirical approach is based on a Roy-type model. Linking the stylized model in Section 3 to the setting described in Brave and Walstrum (2014) and in Carneiro, Heckman, and Vytlacil (2011), I assume that the (log) number of doctor visits is related to the choice of deductible as follows

$$y_i = \delta C_i + \mu(C_i \times \omega_i) + \beta \lambda_i + \psi X_i + \epsilon_i \quad (6)$$

Equation (6) implies that the differences in (log) number of doctor visits $y_i$ associated with choosing a given co-payment rate $C_i$ (a direct function of the deductible level chosen) depend upon an unobserved moral hazard element $\omega_i$, just like in equation (4). $\lambda_i$ captures proxies for healthcare needs, including lagged self-reported health status, reporting a chronic condition, smoking behaviour, index of need for medications in everyday life (0= no needs, 10= very high needs), occurrence of illnesses, accidents or serious health problems in the survey year (*i.e.* health shock), and lagged healthcare utilisation. $X_i$ is a comprehensive set of individual covariates that potentially affect health insurance coverage decisions, including: age, gender, income, household composition, and the canton where respondents were living at the time of the survey. Finally, $\epsilon_i$ is an idiosyncratic error term. As long as $\mu \neq 0$, equation (6) is consistent with the phenomenon that Einav et al. (2013) termed selection on moral hazard. Since any variable that is correlated with the decision to choose a given deductible is also correlated with the unobserved interaction between that decision and moral hazard, I model the selection process explicitly.

The underlying Roy model can be represented in terms of potential outcomes. Without loss of generality, let $D = (0,1)$ indicate whether the individual chose a specific deductible level instead of all others. To simplify matters, let us focus on the case of $D = 1$ indicating the choice of the lowest (highest) possible deductible, so that $D = 0$ indicates the choice of any other higher (lower) deductible. The potential outcomes $y^1$ (doctor visits with the lowest/highest deductible) and $y^0$ (doctor visits for with any other higher/lower deductible) are linearly related to observable need and individual characteristics ($\lambda$ and $X$ respectively), and unobservable components ($\omega^1, \omega^0$), so that

$$\begin{aligned} y^D &= (1-D)y^0 + Dy^1 \\ y^1 &= \alpha^1 + \beta^1 \lambda + \gamma^1 X + \omega^1 \\ y^0 &= \alpha^0 + \beta^0 \lambda + \gamma^0 X + \omega^0 \end{aligned} \quad (7)$$

The deductible choice process is modelled as a latent variable function of observable instruments $Z$ and unobservables $V$ (Heckman, Urzua, and Vytlacil 2006)

$$I = Z\psi - V$$
$$D = \begin{cases} 1 & \text{if } I > 0 \\ 0 & \text{if } I < 0 \end{cases} \quad (8)$$

I use average local group-level premiums and a variable indicating voluntary supplementary health insurance coverage as excluded instruments $Z$ for the selection equation. The average local group-level premium is likely to satisfy the exclusion restriction condition, as (a) premiums are exogenous from the perspective of households (Hadley et al. 2006; Pan, Lei, and Liu 2016), and (b) there is no clear way why the premium should affect the healthcare utilisation response to the health shock directly. Two main arguments support this assumption. Firstly, in the Swiss setting with mandatory health insurance and given a household budget, the most obvious effect of premiums is indirect. Individuals decide on their coverage between September and November of the year prior to reference year. To decide, they acquire information on the premiums on the market, which we encapsulate in the average market premium for the chosen coverage level. Therefore, information on market premiums affects utilisation through coverage choices (type of plan and most notably deductible level) happening at the end of the year before insurance covers for health expenditures. Secondly, any residual influence of monthly premiums on the budget of low-income households, which may prevent them from accessing care to avoid incurring in co-payments - should be contrasted by the redistribution policy based on means-tested subsidies. A similar instrument was used in the context of deductible choices in the Swiss health insurance market by Pichler and Ruffner (2016).

Following Alessie et al. (2020) and Schellhorn (2001), my second instrument is represented by a dummy variable indicating whether the individual has purchased coverage with a voluntary supplementary insurance. Having a comprehensive supplementary insurance allows to cover various (additional or complementary) services that are not part of the basic insurance package. As a result, it is correlated with the choice of deductible although it should not directly affect treatment choices or cost sharing for healthcare covered by the mandatory health insurance, such as doctor visits.

Assuming that the two instruments are uncorrelated with moral hazard, imposing joint normality of the unobserved component with an arbitrary correlation structure $(\omega^0, \omega^1, V) \sim N(0, \Sigma)$, and conditioning on a comprehensive set of explanatory variables $(\lambda, X)$, Brave and Walstrum (2014) and Carneiro, Heckman, and Vyltalcil (2011) show that it is possible to obtain consistent estimates of the marginal treatment effects (MTEs) along the distribution of an unobserved component representing the propensity not to self-select into treatment (in our case, not to choose a given deductible level). The propensity score $P(Z) \in (0,1)$ is estimated with a Probit model. Then, the parametric normal estimator discussed in Brave and Walstrum (2014) is used to correct for selection into deductible and estimate the effect of a given deductible choice (either the lowest or the highest) on healthcare utilisation measured on the intensive margins (log-transformed number of doctor visits). The relationship between MTEs at different values of propensity to not select the deductible and the average treatment effect for the same deductible (which we use as a benchmark) is relatively straightforward. The function recovering MTEs can then be integrated through the estimated propensity score to obtain the average treatment effect (Heckman and Vytlacil 2001). Similar models have been used to estimate structural models allowing for selection on gains across various settings, including universal child care (Cornelissen et al. 2018), higher education (Moffitt 2008), breast cancer treatment (Basu et al. 2007), and health insurance choices (Alessie et al. 2020; Péron and Dormont 2018).

I include a comprehensive set of control variables (see Table 2), capturing the components $\lambda$ and $X$ above. Firstly, I account for habit formation in healthcare utilisation patterns controlling for the lagged value of the dependent variable. Secondly, I account for pre-determined differences in health conditions by controlling for the lagged self-assessed health status, chronic conditions, consumption of medicines, physical activity and smoking status. The combination of initial health status and lagged value of doctor visits provides a reasonable proxy for the expected healthcare need of equation (5). Thirdly, to further balance potential differences in healthcare utilisation, I control for the presence of an unexpected health shocks, which we measure as a new illness, accident, or serious health problem in year. Finally, I include a various socioeconomic and demographic controls that may determine patterns of healthcare utilisation (age, gender, income, receipt of health insurance subsidy, education, and household composition).

To estimate the extent of selection on moral hazard described above in relation to the choice of lowest and highest deductible, my main approach relies on the parametric normal estimator for marginal treatment effects included in the Stata command *mtefe* (Andresen 2018). I model the choice of a health plan with either the lowest (300 Swiss francs) or the highest (2'500 Swiss francs) levels of deductible, separately for years 2018 and 2019, using a Probit link function. At the (second) utilisation stage, I estimate the effect the intensive margin with the log-transformed number of doctor visits. Covariates are included across all model, and are evaluated at their mean when computing MTEs. I report MTE estimates along the distribution of unobserved resistance to select a given deductible, which can be interpreted as the complement of the unobserved level of moral hazard $\omega$, as in equation (4). Standard errors and confidence intervals for MTEs are bootstrapped clustering by canton.

*5.3. Sensitivity and robustness checks*

The parametric normal estimator proposed in Brave and Walstrum (2014) and Andresen (2018) does not strictly require the propensity score to have full common support between treated and untreated groups (compared to alternative semiparametric approaches). To ensure that my findings are not driven by implausible extrapolation over regions with no common support, I re-estimate my models (again separately for 2018 and 2019) comparing the results of parametric normal estimator with analogous models estimated with a semiparametric estimator only defined over regions with common support. For this sensitivity check I use the same covariates and the same excluded variables.

Furthermore, despite the local IV approach isolating plausibly exogenous changes in deductible, my empirical approach may fail to fully account for unobservable circumstances driving both utilisation and deductible choices. Table 3 compares non-switchers and switchers in the 4 groups across a range of individual characteristics for the analytical sample used in this secondary analysis. 81.4 percent of individuals did not change their deductible level between 2018 and 2019. The remaining 18.6 percent are evenly spread across the 4 deductible switching groups introduced in Table 1. There is a substantial balance in observable characteristics across these groups, except for a slightly higher tendency to switch amongst individuals reporting very good health (increasing deductibles), chronic conditions, high need for medications, and not in the labour force (all more likely to stick to their same insurance coverage). These differences are generally consistent with a standard mechanism of adverse selection.

Table 3: Individual characteristics for the 2019 analytical sample across deductible switching groups

|   | *No switching* | | *Strong reduction* | | *Mild reduction* | | *Mild increase* | | *Strong increase* | |
|---|---|---|---|---|---|---|---|---|---|---|
|   | **Mean** | **SD** | **Mean** | **SD** | **Mean** | **SD** | **Mean** | **SD** | **Mean** | **SD** |
| Health status: Very well | 0.168 | | 0.161 | | 0.151 | | 0.196 | | 0.185 | |
| Health status: Well | 0.647 | | 0.684 | | 0.690 | | 0.667 | | 0.699 | |
| Health status: Average | 0.163 | | 0.135 | | 0.138 | | 0.113 | | 0.104 | |
| Health status: Not very well | 0.019 | | 0.019 | | 0.017 | | 0.025 | | 0.012 | |
| Health status: Not well at all | 0.002 | | 0.000 | | 0.004 | | 0.000 | | 0.000 | |
| New illness, accident, diseases | 0.118 | | 0.116 | | 0.103 | | 0.147 | | 0.081 | |
| Smoked in previous year | 0.162 | | 0.155 | | 0.216 | | 0.211 | | 0.185 | |
| Chronic condition in prev. year | 0.467 | | 0.406 | | 0.388 | | 0.392 | | 0.376 | |
| Physical activity in prev. year | 0.840 | 0.367 | 0.813 | 0.391 | 0.858 | 0.350 | 0.843 | 0.365 | 0.879 | 0.328 |
| Medication need in prev. year | 3.156 | 3.591 | 2.335 | 3.082 | 2.720 | 3.453 | 2.711 | 3.385 | 2.023 | 3.155 |
| Woman | 0.553 | | 0.529 | | 0.543 | | 0.578 | | 0.555 | |
| Age | 58.079 | 15.021 | 52.290 | 14.340 | 57.371 | 14.510 | 56.779 | 15.777 | 51.364 | 14.990 |
| Years of education | 14.192 | 3.234 | 14.645 | 3.143 | 14.478 | 3.290 | 14.279 | 3.044 | 14.896 | 3.269 |
| Actively occupied | 0.609 | | 0.748 | | 0.668 | | 0.672 | | 0.740 | |
| Unemployed | 0.011 | | 0.013 | | 0.013 | | 0.015 | | 0.017 | |
| Not in the labour force | 0.380 | | 0.239 | | 0.319 | | 0.314 | | 0.243 | |
| Health insurance subsidy | 0.148 | | 0.174 | | 0.134 | | 0.152 | | 0.069 | |
| Income per household member | 58659 | 41266 | 65066 | 46561 | 65270 | 62000 | 57391 | 32991 | 65316 | 38759 |
| Nr. people in the household | 2.405 | 1.170 | 2.561 | 1.259 | 2.491 | 1.352 | 2.471 | 1.300 | 2.601 | 1.209 |
| Observations | 3352 | | 155 | | 232 | | 204 | | 173 | |

Part of these compositional differences are accounted for by the inclusion of pre-determined health status and other covariates across all models. Moreover, I explore the extent to which some of these characteristics drive my results by conducting stratified analyses across sub-groups defined using these same characteristics (see Section 5.4). However, to further reduce concerns about individual circumstances driving the estimated differences in utilisation and insurance choices, I re-estimated the main models restricting my sample to individuals who – between 2018 and 2019 – reported stable responses for canton of residence (i.e. not moving within the country), health insurance subsidy receipt, self-assessed health status, reporting of a new illness, smoking status, chronic condition, weekly physical activity, needs of medicines in everyday life, working status, and number of people in the household. Descriptive statistics for the group of individuals with stable characteristics are reported in Appendix D.

*5.4. Exploring heterogeneity*

To explore the heterogeneity in the patterns of selection on moral hazard across a number of individual characteristics that may confound the relationship between coverage and utilisation, I conduct a series of stratified analyses. I estimate all stratified models using the parametric normal estimator.

Firstly, albeit the focus on individuals aged 26 and more excludes the behaviour of young adults likely to depend upon their parents, the coverage decisions may nevertheless reflect household decisions and composition (e.g. children) rather than independent individual choices. This would invalidate the empirical approach as it wouldn't allow to directly link heterogeneous responses to a same coverage to a mechanism of selection on an unobserved moral hazard. To check whether household composition confounds the estimates, I propose a set of stratified analyses by household composition (1 member, 2 members, 3 or more members) estimating the Roy model on the full sample pooling the 2018 and 2019 survey waves.

Secondly, risk preferences may be a relevant mechanism triggering changes in health insurance coverage. Unfortunately, detailed questions capturing risk preferences are not consistently recorded in the SHP. However, some evidence suggests that risk and health insurance preferences may vary by gender (Alam, Georgalos, and Rolls 2022; Buchmueller et al. 2013; M. Powell and Ansic 1997). Hence, similarly to the household composition stratification, I estimate my main empirical approach separately for females and males, again on the full sample pooling the 2018 and 2019 survey waves.

Thirdly, one major concern is that individuals' choices may be driven by socioeconomic status rather than explicit or implicit selection based on utilisation responses to health insurance coverage. This pattern would not be fully captured by simply controlling for socioeconomic status in the regression. This is important to consider for two reasons. From a positive perspective, premiums are disjoint from income and subsidies to correct for equity are generally decided ex post, which leaves scope for households to select insurance plans to optimise budget rather than optimal insurance coverage for medical care. From a normative perspective, if low-income households optimise budget rather than coverage, potential changes to policies devised to control the growth in healthcare expenditures in relation to a mechanism of selection on moral hazard may be flawed by unintended distributional consequences. To explore whether estimates of the extent of selection of moral hazard vary across socioeconomic status, I estimate stratified models by receipt of health insurance subsidy and education level pooling 2018 and 2019 waves.

## 6. Results

*6.1. Main results*

My main results are summarized in Table 4. The difference in (log) number of doctor visits between individuals selecting the lowest deductible compared to those selecting any higher deductible (reported as ATE, in column 1) is positive and significant for 2018. The opposite is true for the response to the highest deductible (column 2 for 2018). Assuming that adverse selection based on level of healthcare need is accounted for by the empirical approach, these average differences mainly reflect standard *ex post* moral hazard.

Most importantly, the pattern of MTEs across the distribution of the unobserved resistance to select the corresponding deductible – which I interpret as the reciprocal of the multiplicative moral hazard coefficient (normalised to 1) expressed in Equation (5) - is consistent with the predictions of the model of selection on moral hazard inspired by Einav et al. (2013). For both years, individuals with lowest resistance to select the lowest deductible have a sensibly higher estimated response to coverage compared to the average moral hazard effect. An analogous but inverted pattern is observed in relation to the coverage with the highest deductible: individuals with lowest resistance to select the highest deductible tend to have a sensibly lower response to coverage compared to the average. The response of individuals with higher propensity to select the most/least comprehensive coverage is about 60-80 percent stronger than the average moral hazard effect (ATE), and remains about 25-30 percent stronger contrasting the ATE with the MTE evaluated at the 25$^{th}$ percentile of the unobserved resistance to treatment. This difference in marginal treatment effects is symmetric in magnitude between low and high deductible (i.e. high and low coverage), and roughly similar across the two years analysed. These same results are visually summarised with MTE curves in Appendix C.

Table 4: Results of parametric MTE estimates for coverage
with lowest and highest deductibles in 2018 and 2019

|  | (1) 2018 Lowest deductible | (2) 2018 Highest deductible | (3) 2019 Lowest deductible | (4) 2019 Highest deductible |
|---|---|---|---|---|
| First stage: Propensity score for choice of a given deductible, estimates for instruments | | | | |
| Average market premiums for same coverage | 0.024*** [0.023,0.026] | -0.056*** [-0.061,-0.051] | 0.024*** [0.023,0.026] | -0.048*** [-0.052,-0.044] |
| Covered with supplementary health insurance | -0.195*** [-0.309,-0.080] | -0.075 [-0.296,0.146] | -0.222*** [-0.339,-0.104] | -0.088 [-0.294,0.118] |
| Controls | Yes | Yes | Yes | Yes |
| Second stage: The average difference between individuals selecting the deductible of interest compared to all other deductible levels (ATE) corresponds to the median of MTE distribution | | | | |
| 1$^{st}$ Percentile | 1.087*** [0.599,1.575] | -1.038*** [-1.537,-0.539] | 1.236*** [0.743,1.729] | -1.103*** [-1.753,-0.454] |
| 10$^{th}$ Percentile | 0.759*** [0.514,1.003] | -0.709*** [-0.984,-0.435] | 0.803*** [0.548,1.058] | -0.732*** [-1.091,-0.373] |
| 25$^{th}$ Percentile | 0.568*** [0.448,0.688] | -0.519*** [-0.672,-0.365] | 0.551*** [0.423,0.680] | -0.517*** [-0.712,-0.321] |
| ATE: 50$^{th}$ Percentile | 0.356*** [0.234,0.478] | -0.307*** [-0.402,-0.212] | 0.272*** [0.174,0.369] | -0.277*** [-0.355,-0.200] |
| 75$^{th}$ Percentile | 0.144 [-0.117,0.406] | -0.095 [-0.293,0.104] | -0.008 [-0.236,0.220] | -0.038 [-0.253,0.177] |
| 90$^{th}$ Percentile | -0.047 [-0.449,0.356] | 0.096 [-0.228,0.420] | -0.260 [-0.624,0.105] | 0.178 [-0.202,0.557] |
| 99$^{th}$ Percentile | -0.375 [-1.026,0.276] | 0.425 [-0.127,0.976] | -0.693* [-1.298,-0.088] | 0.549 [-0.122,1.219] |
| N | 4063 | 4063 | 4051 | 4051 |

Notes: Models estimated separately for 2018 and 2019 and for the effect of lowest/highest deductible using the parametric normal estimator for marginal treatment effects included in the *mtefe* stata command. The first stage is estimated with a Probit link function. MTEs are reported at different points in the distribution of U_D, which represents the likelihood not to select a given deductible, where U_D can be interpreted it as $1 - \omega$ with $\omega$ normalized to 1. The selection models use average premium and supplementary insurance coverage as excluded instruments. The utilisation models control for: lagged self-assessed health status, lagged smoking status, lagged chronic conditions, lagged indicator of weekly physical activity, index of medication use in everyday life, gender, age, age squared, receipt of health insurance subsidy, income per person in the household, household composition, years of educations, working status, and cantonal fixed effect. Full regression results in Appendix C, common support graph in Appendix D. 95% confidence intervals were bootstrapped (50 reps) clustering by canton, * $p < 0.05$, ** $p < 0.01$, *** $p < 0.001$.

*6.2. Sensitivity analysis*

The results in Table 4 may mask the role of personal circumstances in driving changes in insurance choices (i.e. deductibles) and/or utilisation. Hence, Figure 3 shows the results obtained using the subsample of individuals with stable personal characteristics as described in section 5.3.

Figure 3: Results for the subsample with stable personal characteristics between 2018 and 2019

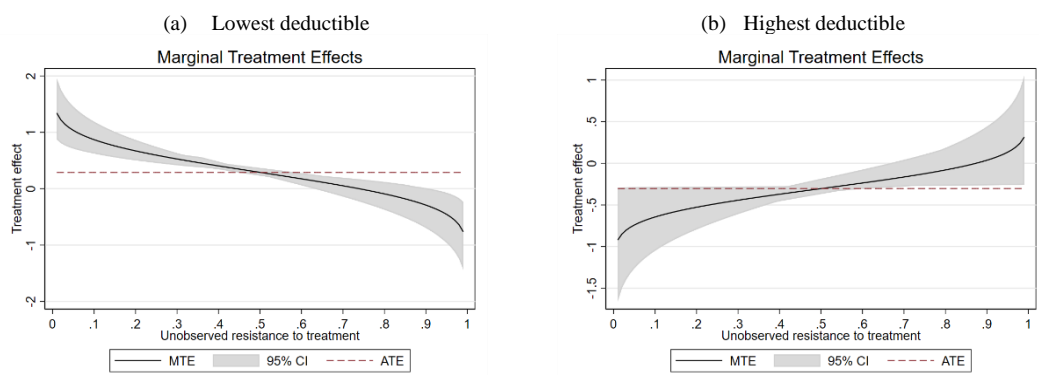

Notes: Models estimated pooling waves 2018 and 2019, separately for lowest and highest deductible levels using the normal estimator for marginal treatment effects included in the *mtefe* stata command. Interpretation of U_D and covariates included are the same as those in Table 4. Full regression results available in Appendix D. 95% confidence intervals (in grey) were bootstrapped (50 replications) clustering by canton.

The result suggests that selection on moral hazard is observed in relation to choosing the lowest deductible level, with a similar magnitude compared to our main results. On the other hand, for the choice of highest deductibles there is no significant difference in the moral hazard estimates across levels of resistance to select the highest deductible. This might suggest that selection on moral hazard is observed for high coverage levels (i.e. low deductibles) conducive to overuse, whilst the decision to purchase a low coverage policy (i.e. high deductibles) is confounded by other factors.

The overlap plots in Appendix D.1 suggest that the common support condition is not satisfied at low levels of resistance to treatment for the results in Table 4. To ensure that my findings are not driven by imprecise extrapolation of my parametric model, in Table 5 I report results obtained using a semiparametric estimator. Importantly, the semiparametric estimator is only defined over regions with common support, hence the comparison only for percentiles 25, 50 and 75 and the difference between ATE and median MTE. In Figure 4 I also report the analogue of Figure 3, but obtained with the semiparametric estimator. A closer comparison between the MTE curves of the two estimators is reported in Appendix D.

Table 5: Results of semiparametric MTE estimates for coverage
with lowest and highest deductibles in 2018 and 2019

|  | (1) 2018 Lowest deductible | (2) 2018 Highest deductible | (3) 2019 Lowest deductible | (4) 2019 Highest deductible |
|---|---|---|---|---|
| Semiparametric estimator | | | | |
| ATE | 0.384*** [0.204,0.565] | -0.297* [-0.567,-0.027] | 0.108 [-0.084,0.300] | -0.218 [-0.479,0.044] |
| 25th Percentile | 0.594*** [0.418,0.769] | -0.449** [-0.734,-0.165] | 0.550*** [0.376,0.724] | -0.521*** [-0.773,-0.269] |
| 50th Percentile | 0.414*** [0.235,0.594] | -0.208 [-0.605,0.189] | 0.358*** [0.163,0.554] | -0.102 [-0.433,0.228] |
| 75th Percentile | 0.085 [-0.151,0.320] | -0.143 [-0.408,0.123] | -0.126 [-0.375,0.124] | -0.105 [-0.349,0.139] |
| Parametric normal estimator (main results in Table 4) | | | | |
| 25th Percentile | 0.568*** [0.448,0.688] | -0.519*** [-0.672,-0.365] | 0.551*** [0.423,0.680] | -0.517*** [-0.712,-0.321] |
| 50th Percentile ATE | 0.356*** [0.234,0.478] | -0.307*** [-0.402,-0.212] | 0.272*** [0.174,0.369] | -0.277*** [-0.355,-0.200] |
| 75th Percentile | 0.144 [-0.117,0.406] | -0.095 [-0.293,0.104] | -0.008 [-0.236,0.220] | -0.038 [-0.253,0.177] |
| N | 4063 | 4063 | 4051 | 4051 |

Notes: Models estimated separately for 2018 and 2019 and for the effect of lowest/highest deductible using the semiparametric estimator for marginal treatment effects included in the *mtefe* stata command. The first stage is estimated with a Probit link function. MTEs are reported at different points in the distribution of U_D, which represents the likelihood not to select a given deductible, where U_D can be interpreted it as $1 - \omega$ with $\omega$ normalized to 1. The selection models use average premium and supplementary insurance coverage as excluded instruments. The utilisation models control for: lagged self-assessed health status, lagged smoking status, lagged chronic conditions, lagged indicator of weekly physical activity, index of medication use in everyday life, gender, age, age squared, receipt of health insurance subsidy, income per person in the household, household composition, years of educations, working status, and cantonal fixed effect. Full regression results in Appendix C. Graphical comparison of results and curves, contrasting parametric normal and semiparametric estimators are included in Appendix D. The estimator is defined only over regions of common support, hence the reason for ATE being different than the median MTE. 95% confidence intervals were bootstrapped (50 reps) clustering by canton, * $p < 0.05$, ** $p < 0.01$, *** $p < 0.001$.

Figure 6: Results obtained with semiparametric estimator

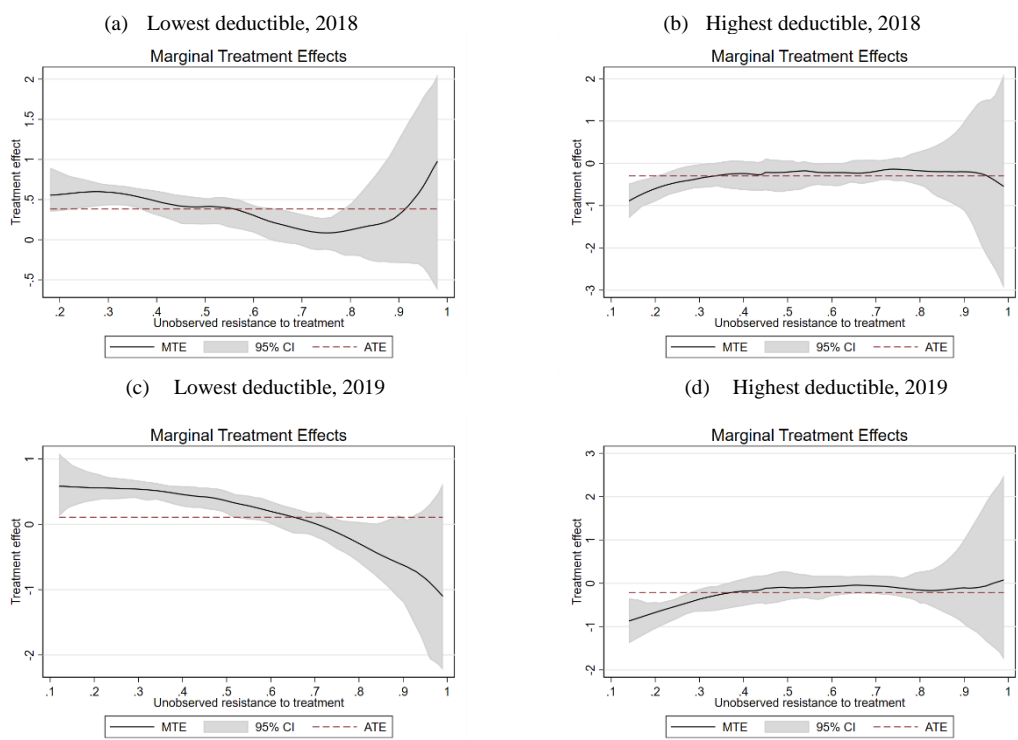

Notes: Models estimated separately for 2018 and 2019 and for the effect of lowest/highest deductible using the semiparametric estimator for marginal treatment effects included in the *mtefe* stata command. The first stage is estimated with a Probit link function. MTEs are reported at different points in the distribution of U_D, which represents the likelihood not to select a given deductible, where U_D can be interpreted it as $1 - \omega$ with $\omega$ normalized to 1. The selection models use average premium and supplementary insurance coverage as excluded instruments. The utilisation models control for: lagged self-assessed health status, lagged smoking status, lagged chronic conditions, lagged indicator of weekly physical activity, index of medication use in everyday life, gender, age, age squared, receipt of health insurance subsidy, income per person in the household, household composition, years of educations, working status, and cantonal fixed effect. Full regression results in Appendix C. Graphical comparison of results and curves, contrasting parametric normal and semiparametric estimators are included in Appendix D. The estimator is defined only over regions of common support, hence the reason for ATE being different than the median MTE. 95% confidence intervals (in grey) were bootstrapped (50 reps) clustering by canton, $^{*}$ $p < 0.05$, $^{**}$ $p < 0.01$, $^{***}$ $p < 0.001$.

In short, the semiparametric models confirm sign and magnitude of the pattern of selection on moral hazard at low resistance to selection of either the lowest or highest deductible level. As shown in Table 5, the magnitude of the MTEs evaluated at the $25^{th}/75^{th}$ percentile of unobserved resistance to treatment is comparable between parametric normal and semiparametric estimators, implying a 25-35 percent difference in responsiveness to treatment compared to average moral hazard. This is indicative of the fact that – whilst the extrapolation beyond the region of common support of the parametric normal model may inflate the magnitude of the heterogeneity in moral hazard at very low levels of resistance to treatment – the findings remain valid and non-negligible in magnitude.

### 6.3. Heterogeneity analyses

Appendix D reports the results of all heterogeneity analyses, namely stratified analyses by household composition, gender, receipt of subsidy, education level. The results of all stratified analyses are largely consistent with the main results. The exceptions are represented by single member households and women. For both these subgroups I find no evidence of heterogeneity in moral hazard in relation to the selection of high deductible plans, whilst the results seem to hold consistent with selection on moral hazard for the choice to purchase low deductible plans. Interestingly, we find no evidence of heterogeneity in responsiveness to coverage for individuals benefitting from premium rebates through a means-tested subsidy for health insurance premiums. It is worth mentioning that the findings of these heterogeneity analyses should be interpreted with caution due to the reduced sample size in some of the sub-samples and the relative need to rely on a parametric approach to estimation.

## 7. Discussion

This paper focuses on selection mechanisms in the intertemporal choice of health insurance deductibles for mandatory health insurance plans in Switzerland, which define the effective level of coverage by limiting the amount of co-payment faced. In contrast with previous published work on Switzerland, the study addresses the notion of selection of moral hazard, that is defined as coverage selection based on heterogeneity in the behavioural response to coverage (Einav et al. 2013), rather than on the level of expected utilisation (or risk). Intuitively, a mechanism of selection on moral hazard is consistent with observing a stronger reaction to high coverage (i.e. low deductible, implying low co-payment) for individuals who previously had lower levels of coverage.

To identify this heterogeneity in moral hazard and selection thereof, I measure utilisation as the number of visits to doctors. I estimate the extent of selection on moral hazard using a structural approach which tries to estimate the difference in healthcare utilisation associated to the choice of lowest or highest deductibles, allowing for selection on unobservables. Building on the stylized model of selection on moral hazard discussed in Section 3, the unobserved selection component can be interpreted as the multiplicative heterogeneous coefficient driving selection on moral hazard. For this Roy-type model, I use average premiums and lagged receipt of health insurance subsidy as excluded instruments in the selection equation. I argue that average premiums are likely to satisfy the exclusion restriction because individuals decide on their coverage each year by the end of November of the previous year, conditional on various covariates. Changes in premiums on the market are exogenous from the perspective of individuals, who may choose lower levels of coverage to benefit from reduced monthly premiums for mandatory health insurance. Hence, the influence of premiums on healthcare utilisation is indirect, through deductible choices. Moreover, given a household budget, the residual effect of the burden of premiums that may push low-income households to refrain from accessing care (i.e. avoid healthcare spending) should be adequately compensated by the means-tested subsidies distributed individually by Cantons through direct reductions of premium bills.

The results of models measuring difference in utilisation across deductible choices accounting for selection on unobservables are consistent with a multiplicative effect of unobserved moral hazard: individuals with highest unobserved moral hazard show differences in utilisation 25 to 35 percent stronger compared to the average moral hazard effect. The sensitivity check on individuals with stable observed characteristics between waves 2018 and 2019 reveal that the results are consistent with selection on moral hazard only for the purchase of high coverage (i.e. low deductible) plans. On the other hand, the selection of plans with the highest deductible level – granting large premium rebates compared to lower deductibles – is likely confounded by changes in personal circumstances, for example changes in income, working conditions, risk preferences or household budget constraints. The analyses exploring the heterogeneity in the results support this idea; I find no significant heterogeneity in moral hazard detected for high deductible plans choice amongst single member households and women in response to high deductible plans. Interestingly, among individuals receiving a subsidy to pay for health insurance premiums I detect no heterogeneity in responsiveness to coverage individuals. These latter findings are consistent with Kaufmann, Schmid and Boes (2017) , who find that individuals receiving subsidies are more likely to purchase high coverage plans but do not show increased their number of GP visits as a results. The same authors also detect differences by gender and household composition, which they attribute to a mix of different risk-taking behaviour, health status, financial constraints, health insurance and financial literacy

Due to differences in the samples, settings and outcome measures the magnitude of my estimates cannot be directly compared to the seminal work by Einav et al. (2013), which substantially inspired my work. Nevertheless, reporting some high-level comparisons is likely informative to gauge the broad plausibility of my findings. Firstly, findings in Einav et al. (2013) imply that average moral hazard is about 28 percent of average healthcare spending in the sample of employees that they analyse (that is 1,330 USD for an average spending of 4,717 USD). My results suggest that the ATE representing average moral hazard is in the same order of magnitude. Remarkably, for the subsample with stable individual characteristics (Figure 3) the point estimate for the utilisation reduction due to average moral hazard is -0.289, hence very close to findings in Einav et al. (2013). Secondly, they propose a comparative statics exercise implying that the average spending reduction per employee increases by roughly 1.4 times when the fraction of employees endogenously selecting the high deductible plans goes from 25 to 75 percent. This increase is essentially due to selection on moral hazard, holding all other characteristics constant (i.e. in absence of selection on moral hazard the endogenous preference for high deductible plans and the resulting average spending reduction would be constant across individuals). An imperfect but close comparison in my setting is given by contrasting the differences at different percentiles of resistance to choosing a given deductible. From Table 5, moving from the 25$^{th}$ to the 75$^{th}$ percentile of unobserved resistance to choosing the lowest/highest deductible implies a change in utilisation response between 1.66 and 1.96 times milder/stronger, hence once again broadly consistent with the findings by Einav et al. (2013).

My study has many relevant limitations. Firstly, it assumes rational individuals ready to change deductible and/or insurer every ear. This assumption fails to consider frictions, inertia and biases which have been highlighted by research in behavioural economics (Frank and Lamiraud 2009). Secondly, the study is based only on two effective time points and on survey data: a longer longitudinal dimension in the data would better address unobserved individual characteristics or underlying healthcare use trends. One relevant timing aspect which these data are silent on is timing of claims, which could inflate observed utilisation rates due to a the form of ex-post adverse selection discussed by Cabral (2017) and Zabrodina (2022). Unfortunately, the data at hand restrict the extent to which I can rule this specific explanation out completely. Thirdly, my study focuses on Switzerland and may have limited external validity. However, the strict standardization of services covered by the mandatory health insurance in Switzerland allows us to focus on the role of the deductible as the main lever available to the insured to reduce the premium paid. Additionally, a country-wide study represents a wider focus compared to most of the published literature trying to address selection on moral hazard. Beyond the policy debate in Switzerland, these findings may also be of interest for countries where a similar insurance-based system has been adopted. Fourth, I use an arguably limited outcome measure, which is self-reported number of doctor visits, focusing on the intensive margin. Access to more detailed claims or administrative data would allow a much deeper understanding of these patterns, a better control for timing of claims, and a more precise assessment of the magnitude of selection on moral hazard in terms of both healthcare utilisation and ultimately costs for the health system. Moreover, focusing on the intensive margin I overlook another crucial decision margin, that is the decision to visit a doctor in the first place. This latter concern is somewhat reduced by the relatively small percentage of people reporting no doctor visits in my sample. Fifth, due to data limitations, I am unable to test the role of other potential relevant explanations for differential responses to health insurance coverage, above all differences in risk attitudes. Finally, the exclusion restriction in the selection equation of the Roy model remains impossible to test formally. The sign of other covariates is intuitively reasonable, providing some reassurance with regards to the validity of the model specification.

Despite important differences in the study settings, my findings are broadly consistent with Einav et al. (2013). In their seminal contribution, they studied health insurance choices of US employees at Alcoa, a large multinational industrial corporation, exploiting an abrupt change in the type of plans offered. Their findings suggested that ignoring selection on moral hazard leads to substantial losses in welfare. My study exploits representative survey data to study patterns of healthcare use and endogenous health insurance plan choice across Switzerland. In a context of steady increase in healthcare expenditures and premiums, if confirmed with more precise individual-level data, my results would call for new ways to regulate the Swiss mandatory health insurance which should incorporate these known features of consumer behaviour in the design of health insurance contracts.

Whilst reforms are most needed on the supply side, on the demand side policy makers may want to explore options to reduce gaming and selection on the side of consumers. Among the solutions discussed in the Swiss health policy arena, a marked increase in the lowest deductible level (set at 150 Swiss francs in 1996, at 230 in 1998 and unchanged at 300 since 2004) or the reduction in the deductible options allowed seem the easiest to implement. To this end, it is interesting how in 2018 the CEO of a leading insurer proposed to increase the minimum deductible to 5'000 or 10'000 Swiss francs a year. Whilst such extreme proposal is unlikely to be politically and socially accepted, the Parliament has discussed and rejected in 2019 an increase in the minimum deductible of 50 Swiss francs per year proposed by the Government[3]. Finding a strong role of selection on moral hazard would imply that these policy levers aimed at fine-tuning deductible with minor changes may turn out to be less effective, compared to what could be expected if selection is based only on different levels of expected healthcare utilisation and projections based on average moral hazard effects. Another alternative is to limit the opportunity to change insurer and/or deductible yearly, either extending the validity of the deductible choice (e.g. setting a multiyear deductible level) or anticipating the deadlines to require a change for the subsequent year. Finally, Einav et al. (2013) postulate the introduction of varying coinsurance rates across diagnoses or types of healthcare, tempering the scope of moral hazard for conditions or treatments with higher discretionary utilisation.

**Appendix**

Available online at https://www.dropbox.com/s/fj0ohwbg17hvuf2/appendix.pdf?dl=0

---

[3] More here: https://www.parliament.ch/en/ratsbetrieb/suche-curia-vista/geschaeft?AffairId=20154157